\documentclass{article}

\usepackage[utf8]{inputenc}

\usepackage{authblk}

\usepackage{algorithm}

\usepackage{algpseudocode}

\usepackage[linguistics]{forest}

\usepackage{color,soul}

\usepackage{hyperref}

\usepackage{float}

\usepackage{ctable} 

\usepackage{algorithm}% http://ctan.org/pkg/algorithm

\usepackage{algpseudocode}% http://ctan.org/pkg/algorithmicx

\usepackage{amsmath,amsfonts,amsthm} % Math packages

\usepackage{graphicx,psfrag,epsf}

\usepackage{enumerate}

\usepackage{mathtools}

\usepackage{natbib}

\usepackage{relsize}

\usepackage{pdfpages}

\usepackage{empheq}

\usepackage{placeins}

\usepackage{lscape}%%%%%pages in landscape mode

\usepackage{forest}

\usepackage{academicons}
\usepackage{xcolor}

\usepackage{academicons}
\definecolor{orcidlogocol}{HTML}{A6CE39}

\usepackage{textcomp}

\usepackage{newtxmath}

\usepackage[normalem]{ulem}

\newcommand*{\N}{{\cal{N}}}

\newcommand*{\bD}{{\mathbf{D}}}

\newcommand*{\bX}{{\mathbf{X}}}

\newcommand*{\bC}{{\mathbf{C}}}

\newcommand*{\bA}{{\mathbf{A}}}

\newcommand*{\bB}{{\mathbf{B}}}

\newcommand*{\calL}{{\cal{L}}}

\newcommand*{\J}{{\cal{J}}}

\newtheorem{assumption}{Assumption}

\title{The policy is always greener: impact heterogeneity of Covid-19 vaccination lotteries in the US.}

\author[1]{Giulio Grossi}

\affil[1]{Department of Statistics, Computer Science, Applications, University of Florence, Italy}

\date{March 2022}

\begin{document}

\maketitle

\begin{abstract}

Covid-19 vaccination has posed crucial challenges to policymakers and health administrations worldwide. In addition to the pressure posed by the pandemic, government administration has to strive against vaccine hesitancy, which seems to be considerably higher concerning previous vaccination rollouts. 

To increase the vaccination protection of the population, Ohio announced a monetary incentive as a lottery for those who decided to vaccinate. This first example was followed by 18 other states, with varying results. 

In this paper, we want to evaluate the effect of such policies within the potential outcome framework, using the penalized synthetic control method. We treat with a panel dataset and estimate causal effects at a disaggregated level in the context of staggered treatment adoption. We focused on policy outcomes at the county, state, and supra-state levels, highlighting differences between counties with different social characteristics and time frames for policy introduction. 
We also studied the nature of the treatment effect to see whether the impact of these monetary incentives was permanent or only temporary, accelerating the vaccination of citizens who would have been vaccinated in any case. 

\end{abstract}


\section{Introduction}

During the Covid-19 pandemic, both scholars and policymakers were challenged by many points of view: in the first emergency, health management focused on containing the pandemic through mainly non-pharmaceutical interventions (NPI from here on), interventions that were undoubtedly effective and decisive in containing the contagions, but unsustainable in the long period. In parallel to health emergency management, research has focused on developing vaccines and treatments against Covid-19.

In particular, with the arrival of the safety and efficacy results of the first vaccines, the organizational plan for the vaccination rollout has begun. However, it soon became clear that the outcome of the vaccination campaign depended not only on the stocks that each country was able to secure but also on the attitude of the population towards vaccination and the policies put in place to facilitate the campaign.

On the one hand, in many countries, there has been part of the population eager to get their vaccine shot, in primis to avoid the most dangerous outcomes of Covid-19 and slow down the spread of the virus, stopping or making his transmission harder. Other parts of the population suspect the fast development of many effective vaccines. A recent stream of thinking refuses optional or compulsory vaccinations, stating that vaccines are not helpful but dangerous for children and adults. Based on fake news or wrong interpretations of scientific results, these arguments have a particular catch, especially in some echo-chambers similar for political orientation and socio-demographic conditions. Nevertheless, more importantly, it could easily harm the effectiveness of the vaccination rollout.

With the spread of more transmissible and pathogenic variants than the original strain of Covid-19 (Alpha, Delta, and most recently Omicron), the time factor has become even more critical in limiting the spread of the disease and avoiding severe consequences for the population, particularly the elderly, residents of nursing homes, and essential workers, such as healthcare personnel. In addition, with the broader availability of vaccines, compliance with the vaccination campaign has become a relevant theme of public health policy. 

Several initiatives have been put in place to entice hesitant to receive the vaccination by states: examples of this type of policies can be monetary incentives to vaccination, or limitations to public life, as has been undertaken by several European states, where certification of vaccination is required to travel by plane or train, go to a restaurant or gym, or even work. Recently, vaccination requirements have also been introduced, limited to the most at-risk population groups in Austria, Greece, and Italy.

The use of monetary incentives, in particular, is of interest, with differences in approach between health policies that used fixed sum incentives (New York City and Pennsylvania in fall 2021) and others, which focused on the use of lotteries. 

This paper focuses on evaluating policies implemented by nineteen US states, which have promoted monetary incentives for vaccination, in the form of lotteries for those vaccinated against Covid-19. The first state to announce this type of policy was Ohio on May 12, 2021, launching the "Vax-a-million" initiative to combat low vaccination levels in the state. Ohio's policy immediately attracted the attention of policymakers in other states, who followed in the subsequent weeks the Ohio example, giving away monetary prizes to vaccinated. At July, 21, 2021, in total eighteen states followed the first example, all except one announce the policy within July, 1, 2021.

Even if, in principle, policymakers design such incentives to help the vaccination rollout, in this specific case, the results are not precise a-priori: on the one hand, we can imagine a positive effect due to the money incentives, coherent with the literature (\cite{campos2021}), but on the other hand, skepticism towards fast-developing vaccines, safety and efficacy doubts and conspiracy theory that flourished around the Covid-19 pandemic, can be enhanced by this kind of public interventions, harming the trust in government (\cite{latkin2021}, \cite{lazarus2021}). Lately, hesitant citizens may value the avoidance of the perceived risk connected to the vaccine more than the probability of winning a lottery prize, see, e.g., \cite{sprengholz2021}.

It turns out to be crucial in such a situation to analyze the outcome of such policies and what drivers are more tightly related to a major or minor impact of these policies. 
Facing this context, assessing the impact of \textit{nudging} toward vaccines is not trivial and could depend on a variety of socio-economic and behavioural factors (\cite{dube2015b}, \cite{savoia2021}, \cite{quinn2016}, \cite{reiter2020}). 

Several papers have investigated the role of incentives in Covid-19 vaccination. Some of them focused on the US (\cite{walkey2021}, \cite{barber2021}, \cite{ab2021}), but none of them, to my best knowledge, have investigated the county level, addressing explicitly for in-states differences in vaccination rollout and incentives effect. At the same time, while the effect of Ohio's program has been studied, there are little to no comparative analyses between US states, while there are theoretical grounds to suspect that different characteristics correspond to different treatment outcomes.

We contribute to the policy evaluation literature by assessing the impact of conditional cash lotteries in a disaggregated framework, in a context of staggered adoption of the policy. We also analyze the duration of the effect, to exploit whether the treatment impact was temporary or persistent. Nevertheless, we aim to focus on the heterogeneity of treatment effects across the counties and identify the socio-demographic characteristics of counties that performed better or worse, filling the gap in the literature concerning states that have undertaken the policy after Ohio. Furthermore, this work contributes to the methodological literature of the synthetic control method by providing estimates of weighted aggregate effects and proposing an inferential procedure for such effects.

The paper is developed as follows: the relevant literature and the context we wish to evaluate are presented in sections \ref{sec:lit} and \ref{sec:disentangle}, Data collection is described in section \ref{sec:data}, while the causal methodology and the inference methods are shown in section \ref{sec:met}, overall results from the analyses are shown and commented in section \ref{sec:res}, section \ref{sec:con} concludes.

\section{Related Literature}\label{sec:lit}

Vaccine hesitancy is a known issue in vaccination rollouts, even before the Covid-19 pandemic, as it was observed in vaccines rollout against measles, HPV, and seasonal influenza, see for a review \cite{dube2013}.

Several health policy interventions in previous years have taken place in order to tackle the concerning trend about reduction of vaccination uptake among children, and in particular, to address directly the parental vaccine hesitancy(\cite{gowda2013}, \cite{williams2014}).

In the majority of previous studies, scholars posed attention to those socio-economic drivers that can explain the variety in vaccination uptakes; see \cite{jarrett2015} for a comprehensive review. 
In particular, \cite{robertson2021}, \cite{razai2021}, \cite{willis2021}, \cite{quinn2016}, \cite{reiter2020}, focuses on the relationship between ethnicity and vaccination uptakes, 
\cite{badr2021} and \cite{azizi2017} shed light on the relation between poverty and unemployment and vaccines, interestingly, before and during the Covid-19 pandemic. 
\cite{bertoncello2020} exploit the inverse correlation between the parental level of education and the vaccine hesitancy and anti-vaccine sentiment, suggesting a similar relationship even when no children are involved. 
\cite{malik2020}, \cite{marks2020} and \cite{joshi2021} investigated the socio-demographic composition of individuals willing to comply in US Covid-19 vaccination campaign, finding out significant differences across ethnicity, gender, and age groups.

\cite{dube2014} highlighted another crucial aspect: vaccine hesitancy is not a "fully-generalized" concept but has several and different drivers across different countries. Therefore every analysis should be exploited to the more granular level available to distinguish drivers between different observation sites.

The unprecedented media coverage about the drug development phase could harm the Covid-19 vaccination rollout. On the one hand, this could have convinced favourable people towards vaccination, underlying the negative relapses of Covid-19. However, on the other hand, the undirected data and information flux have confused many people, leaving a remarkable space for conspiracy threads about the origin of Covid-19 and the development of the three available vaccines in the US. Moreover, the precautionary suspension of viral vector vaccines (Janssen's ad-26 in the US, Astrazeneca's Vaxzevria in Europe) after rare cases of venous thrombosis has generated even more reticence towards vaccines, including those developed using other technologies (mRNA vaccines).

The information source plays a role in determining the attitude towards vaccination: e.g.:\cite{featherstone2019}, \cite{engin2020} and \cite{monsted2022} finds out that vaccine conspiracy belief spreads out on social media, especially among those who express conservative political thought. We can find similar results in Covid-19 vaccine rollout analyses about the US and UK, \cite{loomba2021}.

In this challenging environment, unprecise media coverage (\cite{piltch2021}, \cite{zhou2022}) and live-coverage of phase-3 trial results have confused even more the inhabitants, while regulatory officers (e.g. CDC, WHO, ECDC) have stated on many occasions that approved vaccines were safe and secure.

Effects of these different drivers were heterogeneous among the US, with counties more concerning for the vaccination rollout, especially in the Sunbelt and in the Great plains, with possible profoundly negative impacts also on the Covid-19 cases count and on the related deaths.  

Interesting literature flourished among those incentives for vaccination and health policy interventions that should direct the general population towards health-policy goals, such as reducing smoking, obesity, and alcohol drinking. 
\cite{gorin2015} states that we should not only consider the primary outcome of the policy, which indeed can be distorted by incentives, but also the public discussion that can be generated by the introduction of such policies, increasing the effectiveness of those policies.
\cite{persad2021} analyses from an ethical and legislative point of view the policies undertaken to increase the number of people vaccinated, studying the various dimensions of the positive and negative effects generated by this type of intervention. Positive aspects concern the individual protection against the disease, which sum up to the positive spillover effects for the community, such as the reduction of transmission. The more negative aspects of this policy are the use of public resources to obtain an uncertain outcome and the channelling of a message of vaccine riskiness.
\cite{korn2020} saw vaccination as a kind of social contract, in which the parties which adhered to vaccinating received more empathy from the others who were vaccinated, recognizing the collective use of the policy. In this context, social contrasts between the parties became more acute, reinforcing their positions to the point of partisan behaviour, as highlighted by \cite{weisel2021}.
\cite{campos2021} compared monetary versus educational incentives, finding a 4\% increase in the vaccinated population following a lottery, with minimal increases following an informational incentive. 
\cite{kim2021}  focused on the difference between monetary incentives in the form of a fixed sum and a lottery, noting that lotteries are more effective because, according to the \cite{kahneman2013} perspective theory, individuals fail to correctly assess their chances of winning and are attracted by the jackpot. Another related work from \cite{taber2021} studies whether there are differences between different lottery models, finding no particular differences in the response of individuals. This result is significant for our analysis, as it allows us to compare lotteries across states, even if the amounts dispensed or the probability of winning varied across states. 
\cite{jecker2021} discussed that although monetary incentives can be effective, they are distorting, as poorer people have a greater incentive to participate in politics. Therefore a monetary incentive could be considered unethical, and it conveys the idea that the vaccination choice can be direct. On the other hand, \cite{dotlic2021} argues that the benefit in terms of effectiveness and public policy outcomes is worth the possible distortion that a monetary incentive might create, and however, it is better than coercive measures with negative incentives.

Consequently, we note that there is no agreement in the literature either on the legitimacy of monetary incentives for vaccination or on the actual results of such policies since such incentives may not affect vaccination choices. Nevertheless, governments often used monetary incentives before restricting activities in the absence of a vaccination certificate. 
On the other hand, other countries, such as European countries, saw vaccination as a civic duty and a personal right, relying heavily on the social contract between citizens. Consequently, not provide positive incentives for vaccination but only negative incentives, which had the dual purpose of increasing the number of vaccinated people and limiting the chances of infection for the unvaccinated.

\section{Disentangling treatment effects}\label{sec:disentangle}

Since the announcement by the governor of Ohio, several governors have introduced their vaccine lotteries, creating a natural phenomenon of policy mimicking. While there is some consensus over the positive results of conditional cash lotteries in Ohio (\cite{barber2021}, \cite{acharya2021}); very early evaluations from Arkansas Gov. Hutchinson head in different directions, the lotteries have not been extended because of the population's scarce involvement the policy. 

A substantial problem for policymakers is to assess the heterogeneity of effect after a single treatment. Interaction between pre-existent conditions and treatment effects could explain this kind of variability and could provide helpful insight for future decisions. Unfortunately, early evaluations in several states led to the suspension or rescheduling of the same programs, showing that the effect of the policy could be negative.

We want to investigate which socio-demographic characteristics, besides influencing vaccine hesitancy, also affect the outcome of the policy implemented.

The drivers identified in the literature to explain vaccine hesitancy (see section \ref{sec:lit}) may not be sufficient to explain the performance of different counties. 

A county that is itself highly averse to Covid-19 vaccination might remain so even with monetary incentives to vaccinate, rendering the policy inefficient. 

However, it is theoretically possible that a highly vaccine-averse population will respond very well to the policy because it is susceptible to that kind of incentive, as could be the case with a cash incentive in a low-income population.

\cite{acharya2021} points the same issue in their conclusions: it is dangerous to assume that different counties, environments, and populations react equally to the same interventions, and therefore policymakers should be aware of that. So the question we want to answer is: do the counties that performed better or worse have particular characteristics?

We propose a post-treatment sub-classification of counties according to their reactions to conditional cash lotteries into four groups: 

\begin{itemize}

    \item \textbf{Persistent supporters}: those counties in which incentive has risen the share of vaccinated inhabitants, without a subsequent reduction after the ending of the policy 

    \item \textbf{Anticipators}: Those counties that complied with the policy, with an increase in the share of vaccinated people but later experienced a reduction in the post-treatment period, leading to a realignment to the predicted control value. In these counties, conditional lotteries work as an "anticipator" of the shot, and compliers were inhabitants that could have decided to get the vaccine later but were convinced to anticipate by the policy implementation.

    \item \textbf{Latecomers}: counties in which inhabitants do not get benefit from the policy during the treatment, but after the end of them. In this case, we can state that the policy has a negative but temporary relapse.

    \item \textbf{Persistent opponents}: for these counties, the CCL generates a negative effect on the share of vaccinated people, which endure after the end of the program. This effect may be due to the distrust generated in the population by implementing this type of policy.

\end{itemize}

\section{Data}\label{sec:data}

The dataset used to evaluate lottery incentives for vaccination against Covid-19 includes information on 2925 counties collected from 47 states in the USA. 

The primary outcome of our analysis, the share of 18+ citizens vaccinated against Covid-19, was provided by the Center for Disease Control (CDC). We chose this measure because it is the most responsive to incentives, unlike the percentage of the population fully vaccinated, which instead may show a delay between the adherence to the policy and the baseline outcome due to the time that elapses between the first dose (adherence to the campaign) and the second dose, where the outcome would be measured.

In addition, we decided to focus mainly on the population over 18, both because it is the one that would get the most significant benefits from vaccination against Covid-19. Moreover, at the beginning of the treatment, the authorization for the population over 12 was relatively recent (May 10). Consequently, hesitation behaviours were possible, in addition to the vaccine hesitation measured by the CDC.

The primary outcome is the only observed measure over time, from Jan. 01, 2021, to Aug. 24, 2021. 

We report in Table \ref{tab:des} some descriptive statistics of the main outcome at three-time points: 

\begin{itemize}

\item May 12: Announcement of the first lottery in Ohio.

\item Jul. 01: Announcement of the last lottery in Michigan.

\item Aug. 24: Military vaccination requirement, end of observations

\end{itemize}

While the entire table with the share of first-dose receivers for each state in these three endpoints is presented in table A-2 in the appendix.

\begin{table}[h]

\centering

\footnotesize

\begin{tabular}[t]{|r|c|c|c|}

\hline

  & May 12th & July 1st & August 24th\\

\hline

Mean & 40.220 & 45.677 & 52.282\\

\hline

St.Dev. & 13.439 & 15.985 & 16.680\\

\hline

5\% & 19.681 & 21.055 & 22.101\\

\hline

50\% & 41.674 & 45.228 & 53.627\\

\hline

95\% & 57.785 & 67.690 & 73.630\\

\hline

\end{tabular}

\caption{Descriptive statistics for \% of first-dose receivers}

\label{tab:des}

\end{table}

Alaska, Hawaii, and Puerto Rico were not included in the analysis because of their different characteristics concerning the continental US. Moreover, Texas was omitted because the primary outcome was not collected at the county level. Finally, apart from these four states, not all counties were included due to lack of observations; in particular, we excluded those not reporting vaccinations at the end of the period, on Aug. 24. 

We summarize daily data into weekly data firstly calculating the 7-day moving average of the share of people vaccinated with the first dose. Then we pick the Thursday value as representative of the moving average week.

\begin{figure}[h]

    \centering

    \includegraphics[scale=0.45]{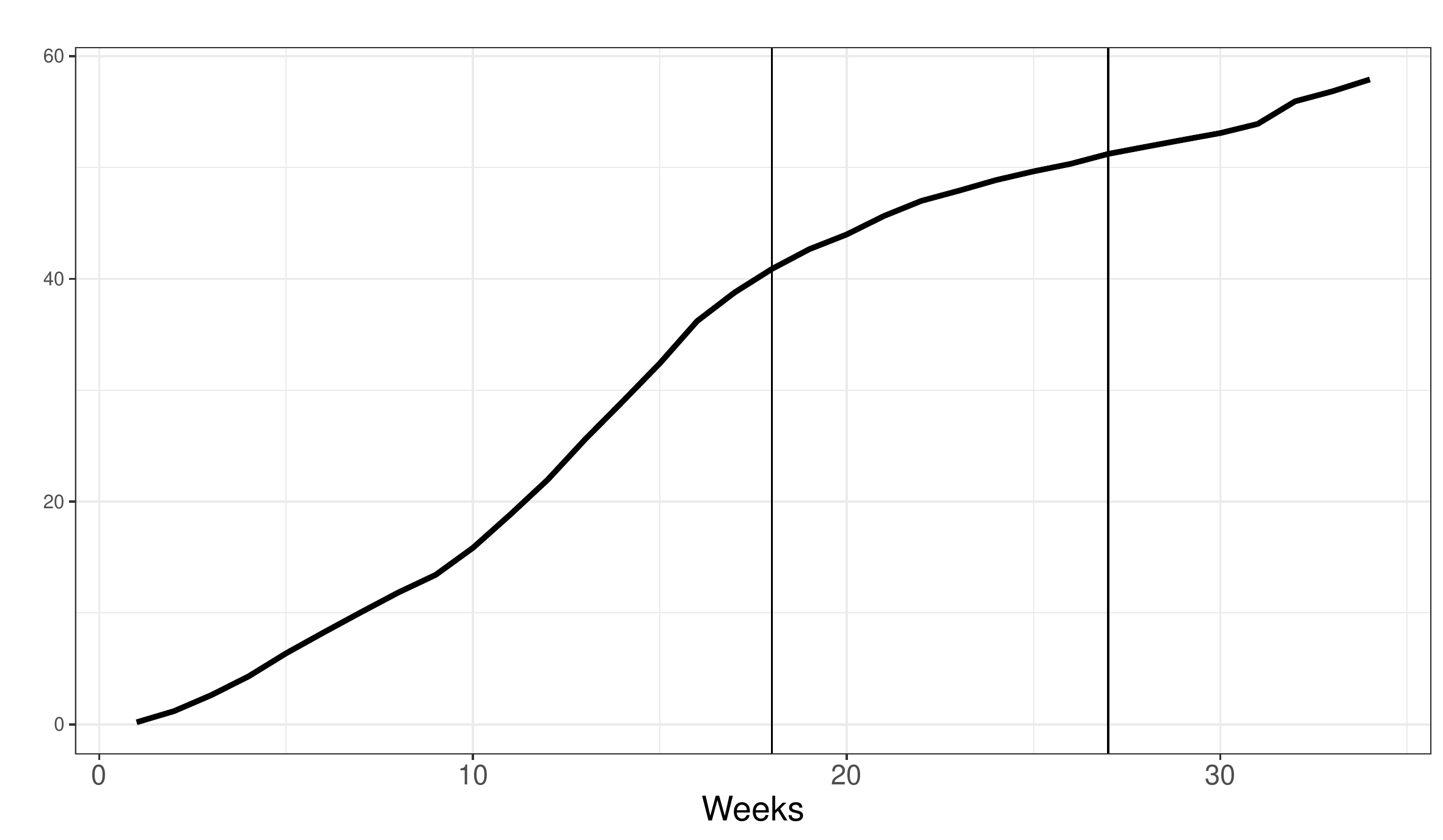}

    \caption{\% of first-dose receivers on the 18+ population - First vertical line: announcement of first lottery (Ohio), Second vertical line: announcement of last lottery (Michigan)}

    \label{fig:fig2}

\end{figure}

The time series of the share of people vaccinated with the first dose is represented in Figure \ref{fig:fig2}, while some relevant statistics of the primary outcome are reported in Table \ref{tab:des}

The dataset we used is enriched by various socio-demographic dimensions that could help us identify the most similar control units concerning the treated units. We choose to add several dimensions as vaccine hesitancy could be driven by many determinants.

In particular, following the literature mentioned above, we choose to focus on those determinants that the field literature has pointed out to determine vaccination compliance. These dimensions are referred to 

\begin{itemize}

    \item Ethnicity

    \item Demographic composition

    \item Political Orientation

    \item Level of education 

    \item Economic indicators 

\end{itemize}

Socio-demographic data were collected in 2020 and provided by the US Census Bureau, as well as environmental characteristics of the counties and the dominant economic activity factors. In addition, the CDC provided data on the percentage of people insured with Medicare. Finally, economic factors, such as the percentage of unemployed and median income by county, are derived from the work of \cite{kirkegaard2016}, updated to 2020. 

The New York Times obtained data regarding voting in the 2020 presidential election.

We also collect from CDC the total number of Covid-19 related deaths. We argue about the importance of this measure since individual behaviours can be altered by a past infection history or a close Covid-19 related death. Therefore, inhabitants of counties in which many fatalities could be more willing to vaccinate themselves. 

Some descriptive statistics for the variables used in the study are shown in Table \ref{tab:predictors}

\begin{table}[h]

\caption{Summary Statistics of socio-demographic characteristics of US counties}

\centering

\footnotesize

\begin{tabular}[t]{|r|c|c|c|c|c|}
\hline
  & Mean & St.Dev. & 5\% & 50\% & 95\%\\
\hline
Hispanic & 9.567 & 13.956 & 0.960 & 4.310 & 40.340\\
\hline
Black & 8.728 & 14.006 & 0.090 & 2.220 & 41.448\\
\hline
Poors & 14.295 & 5.617 & 6.900 & 13.300 & 25.060\\
\hline
Republicans & 65.205 & 15.605 & 35.288 & 68.276 & 85.544\\
\hline
High School & 34.120 & 7.201 & 21.343 & 34.524 & 45.397\\
\hline
College & 22.021 & 9.426 & 11.239 & 19.650 & 41.261\\
\hline
Unemployment & 6.704 & 2.169 & 3.500 & 6.500 & 10.400\\
\hline
Treatment & 38.348 & 48.632 & 0.000 & 0.000 & 100\\
\hline
Deaths/100k & 0.002 & 0.001 & 0.001 & 0.002 & 0.004\\
\hline
Medicare & 11.890 & 4.559 & 5.281 & 11.392 & 19.891\\
\hline
Median Age & 39.912 & 4.794 & 31.700 & 39.900 & 47.800\\
\hline
\end{tabular}

\label{tab:predictors}

\end{table}

\section{Methodology}\label{sec:met}

\subsection{Notation and Setting}

We can consider the evaluation of the conditional lottery policy as a causal inference problem in which some of the units receive the active treatment (the lottery). In contrast, others did not receive any monetary stimulus to participate in the vaccination rollout.

We consider a panel data setting, in which the total set $\N$ of observed units consists in $|\N|=2925$ US counties, observed for $t \in T = (0,\dots, t_{0}, \dots, t_{T})$, $|T|=34$ from the January 1st, 2021 to August, 24th, 2021. On this date, the Pfizer-Biontech vaccine received full approval from the Food \& Drug Administration (FDA starting now). Therefore, a kind of obligation for vaccination was announced for military troops. 
In total, 1134 units have been enrolled in a vaccine lottery in the period considered, while 1791 have not received any kind of monetary incentives for vaccination.
We define the whole set of US counties as $\N$ composed by $n$ units, and specify the set of treated units $i$ as $(i_1, i_2, \dots, i_{M})=\N^{1}$, with cardinality $|M|$, and the set of control units $j$ as $(j_1, j_2, \dots, j_{P})=\N^{0}$, with cardinality $|P|$. Accordingly, we denote as $Y_{n,t}$ the primary outcome, the percentage of residents who received the first dose of vaccine. 

\subsubsection{Treatment Uptake}

We indicate the treatment for unit $n$ in some time point $t$ with $D_{n,t}$ and therefore:

\begin{itemize}

    \item $D_{n,t}=1$ if unit $n$ is receiving the treatment at time $t$

    \item $D_{n,t}=0$, otherwise

\end{itemize}

Following this specification we can construct a \textbf{treatment matrix $D$} as follows:

\vspace{10 pt }

\begin{center}
  $
\textbf{D} =
\begin{bmatrix}\label{mat:tr_mat}

    & i_{1} & i_{2} &  \dots & i_{M}  & j_{1} & j_{2} &  \dots & j_{P}\\ 

0 & 0 & 0 & \dots & 0 & 0 & 0 & \dots & 0\\

1 & 0 & 0 & \dots & 0 & 0 & 0 & \dots & 0\\

\dots & \dots & \dots & \dots & \dots & \dots & \dots & \dots & \dots\\ 

%\hline

t_{0} & 1 & 0 & 0 & 0 & 0 & 0 & \dots & 0\\

t_{1} & 1 & 1 & 0 & 0 & 0 & 0 & \dots & 0\\

\dots & \dots & \dots & \dots & \dots & \dots & \dots & \dots & \dots\\ 

t_{0+h} & 1 & 1 & 1 & 0 & 0 & 0 & \dots & 0\\

t_{T} & 1 & 1 & 1 & 1 & 0 & 0 & \dots & 0 \\

\end{bmatrix}
\vspace{10 pt }
$  

\end{center}

This treatment framework is referred in literature as staggered adoption (\cite{athey2021}, \cite{ben2019}, \cite{callaway2020}), and enables the scholars to evaluate treatment effect occurred in different times. 
In this framework, treated units can receive the treatment at different times, following their entry into the lottery program. In particular, we suppose that no unit receives the treatment if $ t \in (0, t_{0})$. We consider $t_{0}$ as the time in which the first unit receives the treatment, in our case, May 12, in which Ohio Governor announced the "Vax-a-Million" initiative. After time $t_{0}$, treated units $i \in \N^{1} = (i_1, \dots, i_{M})$  can receive treatment at any time, even different in time. On the contrary, the donor pool is composed of $j \in\N^0= (j_1, \dots, j_P)$ units that never receive the treatment.
We found it useful to establish a notation for a unit that will or will not receive treatment at any time $t$: $\bD_n=1$ if $D_{n,t}\neq 0$ for some $t$, and $\bD_n=0$ if $D_{n,t}= 0 \qquad \forall t$.

\begin{figure}[h]

    \centering

    \includegraphics[width=0.75\textwidth]{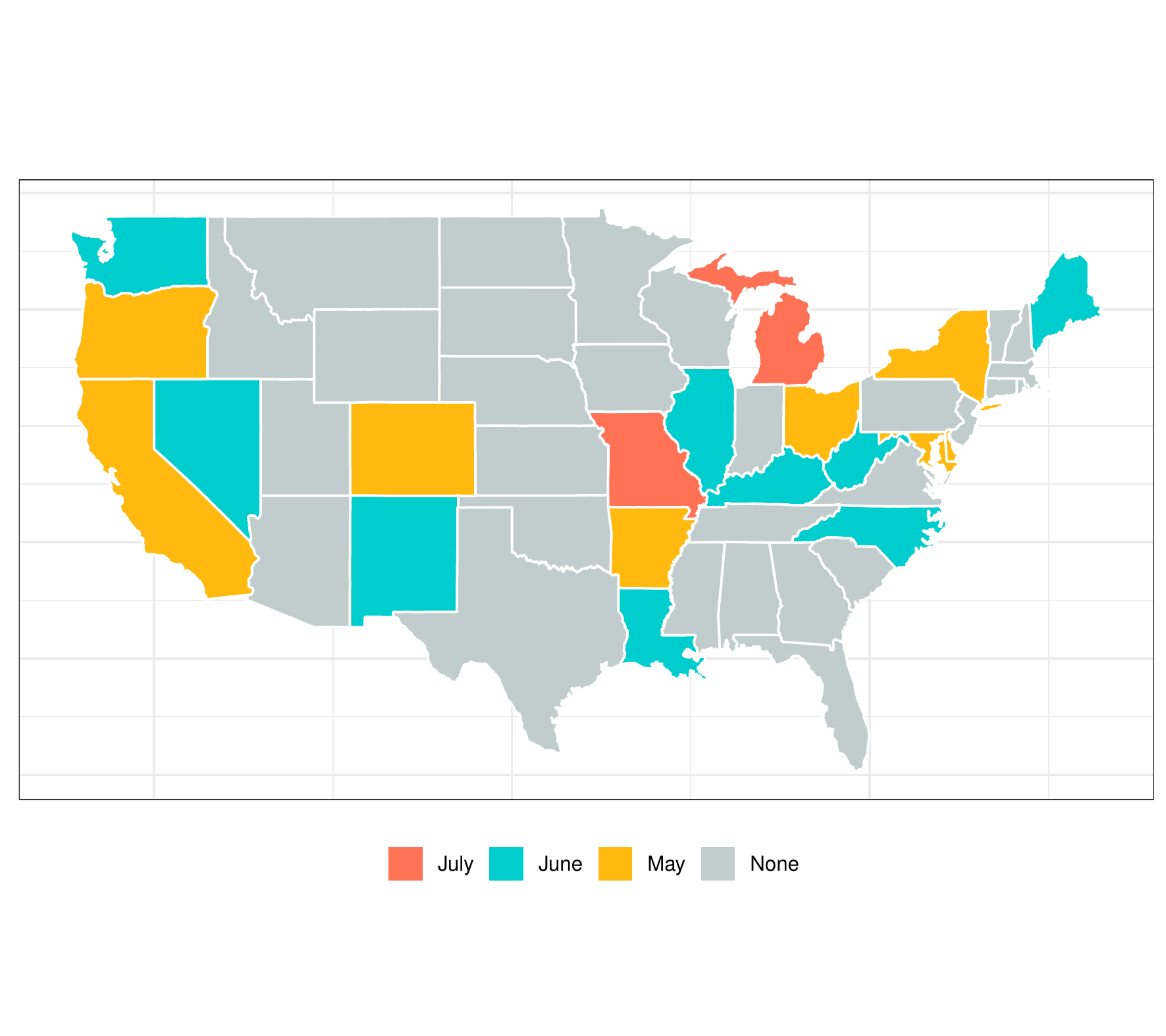}

    \caption{Timing announcement of the vaccine lotteries}

    \label{fig:dannouncement}

  \end{figure}

In total, nineteen states have announced a vaccine lottery to improve the vaccine rollout (Ohio, Oregon, Washington, California, Nevada, New Mexico,  Louisiana, North Carolina, West Virginia, Maine, Kentucky, Michigan, New York, Illinois, Missouri, Arkansas, Colorado, Delaware, Maryland). The beginning time of the lottery and its duration are reported in Figure \ref{fig:dannouncement} and Table A1 in the appendix.

\subsubsection{Assumptions}

In order to define the causal estimand, we wish to state under which assumption we can estimate unbiased causal quantities. Our estimands rely on two basic assumptions: the SUTVA (Stable Unit Treatment Value Assumption) and the no-anticipating treatment assumption.
\begin{assumption}{SUTVA, \cite{rubin1974}}\label{ass:sutva}

\begin{itemize}

    \item Consistency of the treatment

    \item No interference between units

\end{itemize}

\end{assumption}

Consistency means that there is no multiple version of the treatment, which turns out to be challenging in our context. 

It is relatively straightforward to verify that consistency is verified when estimands are calculated at the state level since all counties from the same state receive the same treatment. More significant doubts may arise when comparing or aggregating estimands computed over different states that differ in award amount and probability of winning. We assume that these differences are not relevant for determining whether the treatment causes an effect on our outcome of interest. This assumption appears to be credible, as we find it unlikely that the population applies a quantitative assessment of the cost-benefit ratio associated with vaccination in monetary terms. Consequently, all lotteries are similar for the receiving population. This research question was further investigated by \cite{taber2021}, who found no particular differences in attitudes toward participating in the vaccine lottery among individuals from different states, confirming our assumption.

The non-interference assumption states that the potential outcome $Y_{n,t}$ is a function only of the treatment assignment of the i-th unit and not of any other units, treated or not. We believe that the non-interference assumption may be valid in our context. For a practical example, the fact that a county in Ohio receives treatment should not change the vaccination campaign adherence of an outside county, namely a county in Texas. 
According to this definition for treatment and under assumption \ref{ass:sutva}, we define the potential outcome as $Y_{n,t} = Y_{n,t}(D_{n,t})$ the potential outcome for unit $i$ at time $t$ that we could have observed if the treatment matrix were set to values $D_{n,t} \in (0,1)$, let $Y_{n,t}(D_{n,t}=0) \equiv Y_{n,t}(0)$ and $Y_{n,t}(D_{n,t}=1) \equiv Y_{n,t}(1)$.

We also assume that vaccination compliance is not affected by the expectation about a possible future lottery announcement.

\begin{assumption}{No anticipating treatment}

$$ Pr(Y_{n,t_{-1}}(0)= Pr(Y_{n,t_1}(0)$$ 

$$ t_{-1} \in (0, t_0), t_1 \in (t_{0+1}, T)$$

\end{assumption}

 In theory, people could have changed their behaviour after the lottery announcement, delaying the vaccine administration to get the lottery ticket. 

We can assume this behaviour is not present because most states allowed people to participate in the lottery even though they had already received the first dose. In addition, the short time that elapsed between the announcement of the lottery and the start of the program does not allow for noticeable treatment anticipation phenomena. Lastly, this incentive policy spreads faster among the US, with most treated states announcing the program within 45 days after the original Ohio governor announcement. As a result, we do not expect people to have delayed joining the vaccination campaign to obtain a lottery ticket. 

\subsection{Causal Estimand}

In order to identify the causal effect provoked by the vaccines lotteries in the US, we define as our estimand the quantity. 

\begin{eqnarray}
\tau_{i,t} = Y_{i,t}(1) - Y_{i,t}(0)
\end{eqnarray}

for each county $i \in \N^{1}$ receiving the treatment. We basically compare the observed outcome $Y_{i,t}(1)$ for the treated units, with his counterfactual outcome $Y_{i,t}(0)$, the rate of first dose recipients in the eligible population if the vaccine lotteries were never announced.

Building on this estimand, we are interested in defining causal effects at an aggregate level, defining it as a weighted average of the effects per county $\tau_{i,t}$, multiplied by an appropriate weight $\eta_i$. In our case, we chose $\eta_i$ as the ratio of the population of the i-th county to the total population of the treated set. 

Thus, we can define the weighted  treatment effect in each time point for the pooled treated units as 

\begin{eqnarray}\label{eq:ATT}
\Psi_{\N^{1},t}= \sum_{i=1}^{M} \eta_i \tau_{i,t} \qquad \eta_i \in (0,1)
\end{eqnarray}

Where $\N^{1}$ is the set of considered treated units, with cardinality $|M|$ and $t \in (t_0,T)$, the treatment period.

The definition of $\N^1$ allows us to estimate the effects at the state level, setting $\N^{1}$ equal to the counties that make up the specified state, but also at the supra-state level, by pooling counties from different states (e.g., pooling counties from West Coast). Moreover, when we select the whole set of treated units in all states, we can estimate the overall effect of the policy, pooling the results of all counties treated.

We can also define the overall average effect for the treated set $\N^1$, over the period $(t_1,t_2)$, as 

\begin{eqnarray}\label{eq:PATT}
\Psi_{\N^{1}}= \frac{1}{t_2-t_1}\sum_{t=t_{1}}^{t_2} \Psi_{\N^{1},t}
\end{eqnarray}

The choice of the period $(t_1,t_2)$ allows us to distinguish between the phase in which the lottery was active and the subsequent phase by averaging $\Psi_{\N^{1},t}$, over treatment, or post-treatment periods.

Analysis of the post-treatment period helps us understand whether the positive or negative effect due to the conditional lottery is temporary or permanent. Indeed, it is likely that the lottery serves to convince the portion of the population that we might call latecomers to vaccinate. Thus, lotteries would help to vaccinate faster but not increase the number of vaccinated patients compared with controls at a more distant endpoint. Conversely, if we observed a permanent increase in the number of vaccinated, we could conclude that lotteries affect those who delayed vaccination and those who had no intention of vaccinating. This specification is important since averaging the treatment and post-treatment periods could underestimate the true treatment effect.

Focusing on estimating supra-state weighted average treatment effect, we are posed in a staggered adoption treatment regime. Therefore units in the treated set $\N^1$ did not receive the treatment at the same period, so we have, by definition, a time-imbalanced treatment, with some units for which we observe:

\begin{itemize}

\item $t_0$ pre-treatment periods and $T-t_0$ treatment periods, for the first unit $1 \in \N^1$ receving the treatment

\item $t_{0}+h_{i}$ pre-treatment periods and $T-t_{0}-h_{i}$ treatment periods for the other units $i\neq 1 \in \N^1$

\end{itemize}

For some index $h_{i}$ representing the difference in time points between the first treatment uptake in the treated set $t=t_0$ and the treatment uptake for unit $i$.

We define as pre-treatment period for each unit in $\N^1$ the time span $(h_i, t_0+h_i)$, and as treatment period, namely the latter $t_0$ periods before the treatment uptake, and as treatment period $(t_0+h_i, T-t_0-max(h_i))$ , the firsts $T-t_0-max(h_i)$ periods after the activation of the policy.

We do not focus on the aggregate supra-state post-treatment effect analysis because some units terminated the policy at the last period of the dataset. Consequently, there are no post-treatment periods for some of the units.

\subsection{Penalized SCM }\label{sec:PSCM}

This section explains how potential outcomes not observed are imputed. 

In a context of repeated observations over time, and with many units both under treatment and under control, various tools are possible to assess the effect of the policy under consideration. 

We choose to estimate the causal quantity in equation \ref{eq:ATT} with a modification of the Synthetic Control Method, firstly introduced by \cite{abadie2010}, which is obtaining a growing success among scholars interested in defining causal claims in recent years. 
We estimate causal effects by imputing of missing outcomes, $Y_{i,t}(0)$, namely the outcomes that the treated unit $i$ would have been if it never receives the treatment, constructing a weighted mean of control units in the donor pool. We get a synthetic control that is very close to the treated unit during the pre-treatment period. This method was then modified and extended to allow the estimation of average treatment effects, also in staggered adoption contexts (\cite{dube2015a} \cite{donohue2019}), estimating the treatment effects for each treated unit $i$ and pooling them together.

In particular, among the recent developments of the original estimator (see for example \cite{abadie2021}, \cite{ben2019}, \cite{doudchenko2016}), we choose to adopt the novel methodology developed by \cite{abadie2020}, the so-called Penalized Synthetic Control Method (P-SCM).

We chose this method over the other possibility among the SCM because it is designed to be used in context with disaggregated data, as our framework, and in particular estimates consistent and unique weights, features that are not ensured with the classical SCM.

Following \cite{abadie2003}, we specify the underlying data structure for the principle outcome, as :

\begin{eqnarray}
Y_{n,t}= \bA_t + \sum_{o=1}^O\bB\zeta_o + \sum_{u=1}^U \textbf{R} \phi_u  + \epsilon_{n,t}
\end{eqnarray}

So the outcome variable $Y_{n,t}$ follows the typical specification of a linear factor model, widely used in the SCM literature. The outcome is defined by the linear combination of the common time-trend $\bA_t$, $O$  observed factors $\zeta$, and $U$ unobserved factors $\phi$, with $\bB, \textbf{R}$ as the vectors of factor coefficients.

We define the estimation of the missing outcome ${Y}_{i,t}(0)$ as 

\begin{eqnarray}
\widehat{Y}_{i,t}(0) = \sum_{j=1}^{P}  \omega_{j} Y_{j,t}  \qquad t \in (t_{0}, \ldots, T)
\end{eqnarray}

We define $\bX_{i}=\left[Y_{i,0}, \dots, Y_{i,t_0} \right]'$ as a $t_0 + h_i-$dimensional vector of pre-treatment outcomes,for each treated unit $i$, and given a positive penalization constant $\lambda^{(i)}$, $i \in \N^{1}$ , the set of weights 

$$\boldsymbol{\omega}^{(i)} = \left[ \{\omega^{(i)}_{1}, \ldots,
 \omega^{(i)}_{j}, \ldots,\omega^{(i)}_{{P
 }} \right]'$$ 
 defines
the \textit{penalized synthetic control unit} of unit $i$. While the set of weights $\boldsymbol{\omega}^{(i)}$ is the solution to the following minimization problem in equation \ref{eq:hatweights}.

\begin{eqnarray}\label{eq:hatweights}
 \arg \min_{\boldsymbol{\omega}^{(i)} \in \boldsymbol{\Omega}}
\left \| \bX_{i}  - \sum_{j  \in \N^0}\bX_{j}\omega_{j}^{(i)}\right \|^2 + \lambda^{(i)}
\sum_{j\in \N^0} \left \| \bX_{i}  - \bX_{j}  \right\|^2
\end{eqnarray}

  subject to $$\omega^{(i)}_{j}\geq 0 \quad \forall j \in \N^0;\qquad \sum_{j \in \N^0} \omega^{(i)}_{j}=1,$$

  with $\|\cdot \|$ is the $L^2-$norm: $\| \mathbf{v} \| = \sqrt{\mathbf{v}' \mathbf{v}}$ for   $\mathbf{v} \in \mathbb{R}^r$

Technical details of the estimator and its use can be found in \cite{abadie2020} and \cite{abadie2021}.

The main reason to choose this model over the different possibilities is the specific design of the P-SCM estimator for estimating average treatment effects with disaggregated treated units, granting us the uniqueness of the weights. 

Moreover, it enables us to maintain an agnostic behaviour towards the choice between matching methods and SCM, leaving the choice to a cross-validation procedure on the donor pool. As discussed in \cite{athey2018}, in observational studies where both N and T are large, the choice of which method to use is not straightforward, and in our opinion, P-SCM may allow a data-driven solution to this problem. In fact, when $\lambda^{(i)} \to 0$, the P-SCM collapses into the standard SCM, while when $\lambda^{(i)} \to \infty$ the P-SCM is equivalent to the nearest-neighbor matching estimator.

The tuning parameter $\lambda^{(i)}$ were chosen by using a weighted cross-validation approach, derived from \cite{abadie2020}: we define the function $\Phi(\lambda)$, exposed in Equation \ref{eq:phi}, which minimizes the overall Root Mean Square Prediction Error $\Re(\lambda, T-t_{0})$ of the donor pool for each level of $\lambda$ in the treatment period. 

$$
\Re(\lambda, T-t_{0} )=\sqrt{\dfrac{1}{(T-t_{0})}\sum_{t={t_0}}^{T} \left[ Y_{j,t}-\widehat{Y}_{j,t}(\lambda)\right]^2}
$$

\begin{eqnarray}\label{eq:phi}
 \lambda^* = \arg \min_{\lambda \in \Lambda}\Phi(\lambda)=\mathlarger{\sum}_{j=1}^{\N_j}\xi_j\times \Re(\lambda, T-t_{0} )
\end{eqnarray}

with $$\xi_j\in (0,1); \sum_{j=1}^{P}\xi_j=1$$

$\Lambda$ is the set of possible values for $\lambda$, 100 values ranging from 0 to 1, each 0.01. $\xi_j$ is a weight representing the relative importance of the control unit $j$, calculated as

$$ \xi_j=\frac{\pi_j}{\sum_{j=1}^{P}\pi_j}$$ 

Where $\pi_j$ is the propensity score for control unit $j$ , described in section \ref{sec:prscore}. Values of lambda used in the estimation are reported in table A4 of the Appendix.

\subsubsection{Donor pool definition}\label{sec:prscore}

In order to define the effect of conditional cash lotteries on vaccination rates in treated US counties properly, we choose to select the donor pool by using the propensity score approach, firstly introduced by \cite{rosenbaum1983}. Formally we can define it as the probability of being assigned to the treatment at any time ($\bD_n=1$), or control ($\bD_n=0$), given a particular set of covariates $\bC$ (see Table \ref{tab:predictors}). 

\begin{eqnarray}
\pi_n=\pi_n(\bC)=Pr(\bD_n=1|\bC)
\end{eqnarray}

under the assumption that 

\begin{eqnarray}
Pr(\bD_1, \dots, \bD_n | C_1, \dots, C_n) =
\prod_{n=1}^{N}\pi(\bC)^{I(\bD_n=1)}(1-\pi(\bC)^{1-I(\bD_n=1)})
\end{eqnarray}

The propensity scores should help us obtain a homogeneous donor pool in characteristics concerning the treatment pool, allowing us to estimate causal effects from a "like with like" comparison.

In practice, we estimate the propensity score on a set of covariates describing socio-economic, political, ethnic, and demographic characteristics (see Table \ref{tab:des} for the complete list).

Once we have estimated $\pi=\{\pi_j\}_{j=1}^P$ for the control units, we select the donor pool by choosing the units with $\pi_j\geq med(\pi)$. Finally, two-sample t-tests to evaluate the similarity between covariates of the treated pool and donor pool for each state are reported in Table A3 in the appendix.

Such a procedure for donor pool restriction speeds up weights computations. Moreover, it allows to build the synthetic control of the treated unit from similar control units, without relevant losses in terms of prediction error, see also \cite{abadie2010}.
In addition, calculating the propensity score as a treatment likelihood measure can be helpful to treat differently the control observations that could have been more likely to be assigned to the treatment. 

Specifically, we use propensity scores as weights in both the definition of the hyperparameter $\lambda^*$  in section \ref{sec:PSCM} and in the definition of the p-values for the causal effect, in section \ref{sec:inference}.

\subsubsection{Inference}\label{sec:inference}

Although defining confidence intervals and p-values for causal effects obtained using synthetic control methods is not straightforward, many scholars have proposed different methods to assess the significance of such causal effects.

The seminal idea for inference in such contexts was proposed by \cite{fisher1936}, based on randomization inference, and subsequently \cite{abadie2010} and \cite{abadie2021} proposed a pseudo p-value for causal effects.

This approach aims to reassign the treatment to control units in the donor pool and estimate the so-called placebo effects in this way. Therefore, we can compare the estimated treatment effect with the distribution of placebo treatment effects, considering a treatment effect significant when the magnitude of the effect is large concerning the distribution of the placebo effect. As noticed by \cite{abadie2010}, this approach should be taken carefully because it can be biased when there is a poor pre-treatment fit of counterfactuals. \cite{abadie2010} introduces a test statistic that can make the effect for the treated unit comparable with the placebo effects estimated for the control units. For this purpose, it is defined a test statistics $\theta_n$ as 

\begin{eqnarray}\label{eq:theta}
\theta_n=\frac{{\Re}(t_{0},T)}{\Re(0,t_{0})}
\end{eqnarray}

Where $\Re(t_1,t_2)$ is the root-mean square prediction error calculated between $t_1$ and $t_2$ as 

$$ \Re(t_1,t_2)=\sqrt{\frac{1}{t_2-t_1}\sum_{t=t_1}^{t_2} \tau_{n,t} }$$

Consequently, $\theta_n$ represents the ratio of the mean prediction error between treatment or post-treatment period, and pre-treatment. In this way, we can compare units that have different fitting measures, and thus directly compare $\Theta_j = \{\theta_j\}_{j=1}^{P}$, with the test statistic for the treated unit $\theta_i$.

Therefore, under an exact null hypothesis $\Psi_{\N^1}=\Psi_{\N^0} \quad \forall j\in \N^0, \forall t \in T$ we can define the p-value as:

\begin{eqnarray}\label{eq:abadiep}
p_i= \frac{\sum_{j=1}^{P}\vmathbb{I}(\theta_i>\theta_j)}{|P|}
\end{eqnarray}

We reject the null hypothesis of no effect whatsoever if $p_i \leq \alpha$, with $\alpha$ some pre-specified significance level, e.g. $\alpha=0.1$.

 $\vmathbb{I}$ is the indicator function, counting the number of times in which some control units' test statistics exceed the treated unit $i$ test statistics.

Such procedure assumes that each control unit has the same probability of being assigned to the treatment, and therefore, they have equal weights in determining the p-value, namely $\frac{1}{|P|}$.

A recent proposal from \cite{firpo2018} slightly modifies the p-value expressed in equation \ref{eq:abadiep}, allowing for weights that should distinguish between units that are more and less likely to be assigned to the treatment.

\begin{eqnarray}\label{eq:firpop}
\rho_i= \sum_{j=1}^{|P|} \kappa_j \times \vmathbb{I}(\theta_i>\theta_j)
\end{eqnarray}

In principle, it could not be easy to find such weights $\kappa_j$, but in our context, we can use as a proxy of them the estimates for the propensity score $\pi_j$, defined in section \ref{sec:prscore}. 

In this study, we focus on assessing the significance of the aggregate $\Psi_{\N^1}$ effects, and, as a result, we do not estimate p-values or confidence intervals for individual effects for each i-th county. 

In order to evaluate the aggregate causal effect for each US state or particular subsets of the treated pool, we construct many placebo sets for each treated set, adapting the inference routine described by \cite{cavallo2013} for the inference using SCM in aggregate frameworks.

Our procedure for inference is illustrated by the algorithm \ref{alg:inference}.

\begin{algorithm}[H]\label{alg:infpvalue}

\caption{SCM inference with many treated units}\label{alg:inference}

\begin{algorithmic}

\Procedure{p-values}{$\tau_j(\bD_j=1)$, $\Theta_{\N^1}$, $\pi_j$}

\Require $\tau_j(\bD_j=1)$, $\Theta_{\N^1}$, $\pi_j$ 

\Ensure $\rho_{\N^1}$

\begin{enumerate}

      \For{$\J \in (1:1000)$}

    \item Sample with replacement $M$ units from the donor pool $\N^0$, forming the placebo state $\J$

    \item Calculate $\Psi_{\J,t}$ and $\Psi_{\J}$ according to equation \ref{eq:ATT} and \ref{eq:PATT}

    \item Calculate $\Theta_{\J}$ for each placebo state as in equation \ref{eq:theta}

    \item Calculate 

    $$\calL_{\J}(\pi_j) = \prod_{j=1}^{M}\pi_j$$ 

    as the likelihood of observing $\bD_j=1$ for the units composing the placebo state $\J$

  \EndFor

          \item Calculate the p-value $\rho_{\N^1}$ for the treated state by using Equation

          $$ \rho_{\N^1}=\mathlarger{\sum_{\J}}\frac{\calL_{\J}}{\sum_{\J}\calL_{\J}} \times  \vmathbb{I}(\Theta_{\N^1}>\Theta_{\J}) $$

\EndProcedure          

\end{enumerate}

\end{algorithmic}

\end{algorithm}

From the distribution of effects for the placebo states $\Psi_{\J,t}$, we can also derive the quantiles of the distribution to be compared with the treatment effect $\Psi_{\N^1,t}$. 

Moreover, to assess the significance of causal impact for each state more precisely, we have distinguished between the moment in which the policy was in charge and the moment in which the lottery ended.

\subsection{Units Clustering}\label{sec:cluster}

In this section, we aim to identify cluster of counties to describe the possible differences in treatment effects. The only vaccine hesitancy could not be enough to explain the negative results of conditional cash lotteries, and other patterns could be exploited in order to provide insights for future health policy decisions.

For this purpose, we decided to classify treated counties according to invariant characteristics of the same, using variables socio-demographic, economic, and cultural character. The classification of counties was performed following a clustering approach with Gaussian mixture models, see for details \cite{fraley2002}, \cite{mclachlan2019}. We choose the number of clusters that minimize the selection model's BIC for membership in each group of counties through this data-driven procedure. According to the set of covariates $\bC$, we found that six clusters of counties are identified, the mean values of the predictors are reported in the Table \ref{tab:clustdes}. 

\begin{table}[H]

\centering

\footnotesize

\resizebox{\textwidth}{!}{%

\begin{tabular}[t]{|r|c|c|c|c|c|c|}

\hline

Cluster & 1 & 2 & 3 & 4 & 5 & 6\\

\hline

Hispanic & 4.594 & 18.005 & 5.151 & 50.235 & 9.831 & 3.682\\

\hline

Black & 4.554 & 8.001 & 36.641 & 2.351 & 2.431 & 2.729\\

\hline

Poors & 14.402 & 8.206 & 20.699 & 16.966 & 12.640 & 14.461\\

\hline

Republicans & 64.807 & 40.393 & 53.842 & 52.314 & 56.809 & 71.797\\

\hline

High school & 36.100 & 21.935 & 36.114 & 28.599 & 27.980 & 36.533\\

\hline

College & 21.496 & 40.133 & 19.106 & 19.042 & 26.600 & 17.154\\

\hline

Unemployment rate & 7.720 & 7.698 & 7.880 & 9.541 & 8.550 & 7.576\\

\hline

Deaths/100k & 153.022 & 121.300 & 235.175 & 174.104 & 99.423 & 293.646\\

\hline

Medicare & 10.112 & 8.277 & 11.019 & 8.987 & 17.856 & 14.809\\

\hline

Earnings & 25475.601 & 37966.388 & 24539.974 & 23472.680 & 25117.741 & 24344.540\\

\hline

Median Age & 39.307 & 37.402 & 37.813 & 35.257 & 46.128 & 41.918\\

\hline

\end{tabular}

}

\caption{Mean covariates per cluster}

\label{tab:clustdes}

\end{table}

Looking at the predictors, we can see that clusters 3 and 4 have a very different ethnic composition compared to the others, particularly in cluster 3, the black population is relevant, while cluster 4 represents the one with the highest percentage of Hispanics. Cluster 2 has the richest, the most educated population on average and the lowest level of voting for the Republican Party. Clusters 1 and 6 are distinguished by their high percentage of votes for Republicans, but in cluster 6, the Covid-19 provoked many more deaths with respect to the other clusters. Cluster 5 has the highest median age, and the most share of citizens ensured with Medicare.

\section{Results}\label{sec:res}

\subsection{Causal effects}

The estimation with the matching+P-SCM gives us a reliable picture of the results of conditional cash lotteries used to promote vaccination in the US. 

In Figure \ref{fig:fig4} is reported the  average treatment effect for each treated county, in particular, we can underline over-border positive results in the north-west area (Oregon, Washington, and some areas of California), in the Midwest (Ohio, Kentucky, Illinois, West Virginia), and the New England (New York, Maryland, Delaware). In the other areas, the effect of the lotteries does not seem to be relevant, and in particular minor or adverse effects can be seen in Sun Belt states. Moreover, relevant differences can be found inside each state, suggesting an interaction between counties characteristics and treatment effects.

At state-level treatment effects, we notice a positive and significant impact for Ohio, West Virginia, Oregon, Washington, Kentucky, Nevada and Illinois, while we observe adverse and significant effects for Louisiana. However, the positive treatment effects turn out to be greater in magnitude and more widespread concerning the negative effects. We do not notice significant results for the other states. 

In post-treatment, however, we see that many states with positive results do not consolidate the acquired margin.
Thus, vaccination levels return to those expected from units in the donor pool; examples are the time series of the effect in Ohio, Illinois, Kentucky and Nevada.
Therefore, we should conclude that the effect of the policy was in general temporary, except the West Virginia, Washington and Oregon, which continue to show a significant and positive effect during the post-treatment period.  We excluded Missouri from the results, as the adoption of the policy is too late compared to the others (Jul. 21).

\begin{figure}[ht]

    \centering

\includegraphics[scale=0.75]
    {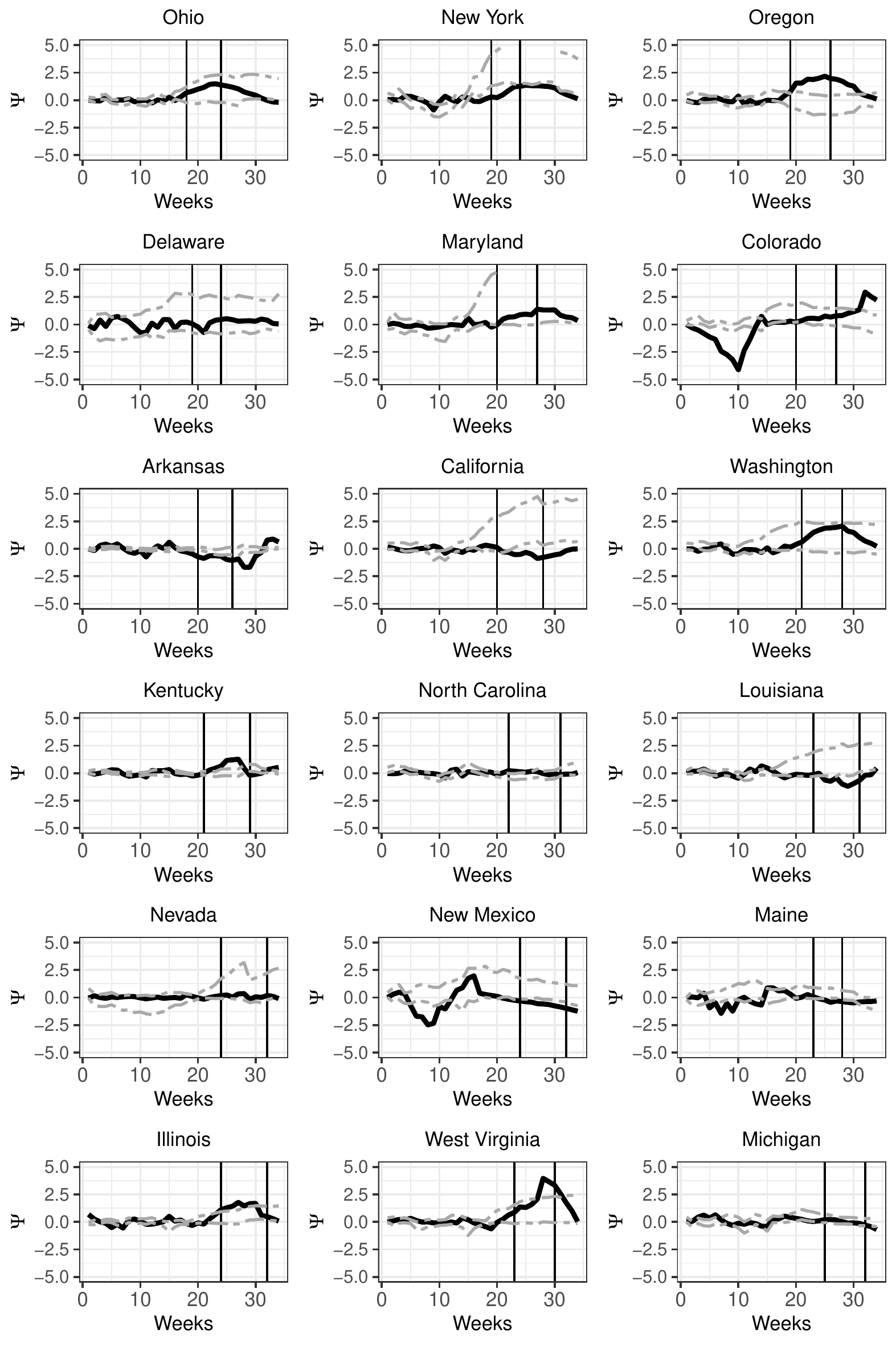}
    \caption{Intertemporal effect $\Psi_{\N_i}$}
    \label{fig:fig3}

\end{figure}

\begin{figure}[h]

    \centering

    \includegraphics[width=\linewidth]{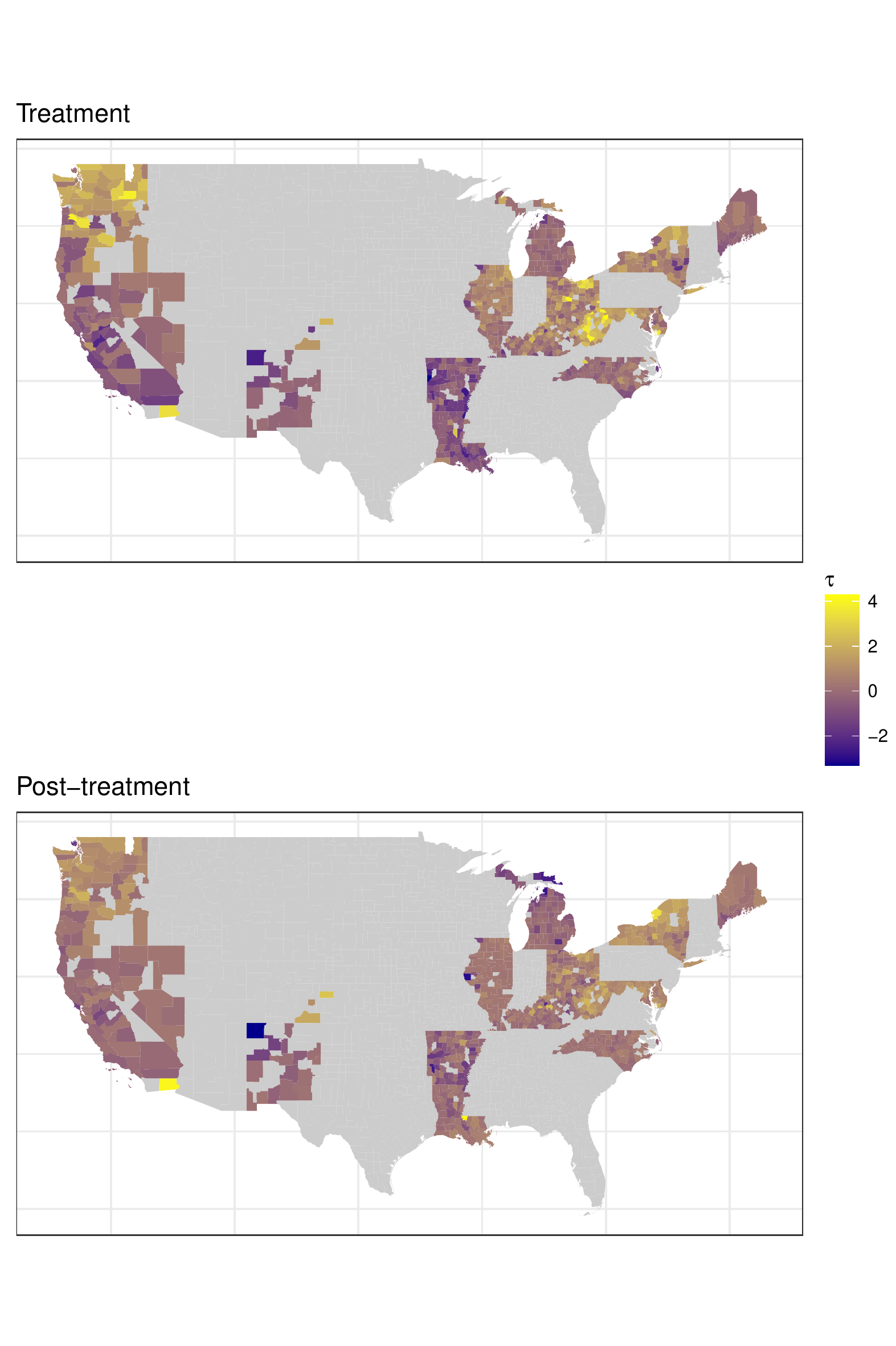}

    \caption{treatment and post-treatment mean effect at county level}

    \label{fig:fig4}

\end{figure}

The full results at the state level, with the estimated effect of the treatment and the p-value for each state, are shown in the Table \ref{tab:results}, the time series of the treatment effect with the 5\% and 95\% quantiles of the distribution of placebo states are shown in the Figure \ref{fig:fig3}. 

In general, the root mean square prediction error shows an error of less than one percentage point, which can be considered an acceptable result. However, results from states with errors close to or exceeding this level (Colorado and New Mexico) should be treated with caution.

\begin{table}[h]

\label{tab:results}

\centering

\footnotesize

\begin{tabular}[t]{|r|c|c|c|}

\hline

  & RMSPE & Treatment & Post-treatment\\

\hline

Ohio & 0.283 & 1.208 & 0.504\\

%\hline

p-value &  & 0.000 & 0.392\\

\hline

New York & 0.431 & 0.806 & 0.930\\

%\hline

p-value &  & 0.192 & 0.186\\

\hline

Oregon & 0.391 & 1.855 & 0.967\\

%\hline

p-value &  & 0.000 & 0.020\\

\hline

Delaware & 0.513 & 0.001 & 0.323\\

%\hline

p-value &  & 0.376 & 0.859\\

\hline

Maryland & 0.324 & 0.847 & 0.920\\

%\hline

p-value &  & 0.130 & 0.060\\

\hline

Colorado & 1.806 & 0.598 & 1.716\\

%\hline

p-value &  & 0.976 & 0.249\\

\hline

Arkansas & 0.486 & -0.806 & -0.369\\

%\hline

p-value &  & 0.509 & 0.004\\

\hline

California & 0.358 & -0.513 & -0.315\\

%\hline

p-value &  & 0.116 & 0.991\\

\hline

Washington & 0.650 & 1.717 & 0.927\\

%\hline

p-value &  & 0.000 & 0.031\\

\hline

Kentucky & 0.389 & 0.657 & 0.229\\

%\hline

p-value &  & 0.000 & 0.959\\

\hline

North Carolina & 0.306 & 0.036 & -0.040\\

%\hline

p-value &  & 0.037 & 0.999\\

\hline

Louisiana & 0.319 & -0.699 & 0.039\\

%\hline

p-value &  & 0.002 & 1.000\\

\hline

Nevada & 0.143 & 0.136 & 0.006\\

%\hline

p-value &  & 0.042 & 0.960\\

\hline

New Mexico & 1.621 & -0.611 & -1.182\\

%\hline

p-value &  & 0.848 & 0.020\\

\hline

Maine & 0.799 & -0.292 & -0.397\\

%\hline

p-value &  & 0.997 & 0.972\\

\hline

Illinois & 0.461 & 1.270 & 0.179\\

%\hline

p-value &  & 0.000 & 1.000\\

\hline

West Virginia & 0.416 & 2.466 & 1.338\\

%\hline

p-value &  & 0.000 & 0.001\\

\hline

Michigan & 0.438 & -0.003 & -0.565\\

%\hline

p-value &  & 0.102 & 0.000\\

\hline

\end{tabular}

\caption{State-level effects estimation}

\end{table}

\subsection{Staggered adoption results}

In addition to the causal effects for each state, we wanted to assess the aggregate effects for several states that began vaccine lotteries simultaneously, or at least within a short time interval.

 The hypothesis we want to test is that the first introduction of the policy gives some novelty effect. As a result, the early states to have introduced the policy will have better outcomes than the laggards. This idea, in particular, would be explained by the considerable media buzz coming from the Ohio announcement, mirrored in earlier studies that have addressed the assessment of this policy, focusing only on the results of the first state.

In order to evaluate this type of effect, we want to test different groups of states, subdivided according to their entry into the policy. In particular, we have defined the following groups:

\begin{itemize}

\item Early Bird: adoption within May 20 (Ohio, New York, Oregon, Delaware)

\item Second Echelon: adoption by the end of May (Maryland, California, Arkansas)

\item Third Echelon: adoption within mid-June (Washington, Kentucky, North Carolina)

\item Latecomers: adoption by the end of June (Louisiana, Nevada, Maine, Illinois, West Virginia, Michigan)

\end{itemize}

In addition to this classification, we divided the pool of all treaties into two sections: states that announced the vaccine lottery in May and those in June; Michigan, which announced it in July, 01, was included in the June pool. We excluded Colorado and New Mexico counties from this analysis due to the unreliability of the pre-treatment fit. 

\begin{table}[h]

\centering

\footnotesize

\begin{tabular}[t]{|r|c|c|}
\hline
  & ATE & p-values\\
\hline
Adoption in Early May & 1.075 & 0.002\\
\hline
Adoption in Late May & -0.260 & 0.973\\
\hline
Adoption in Early June & 0.843 & 0.003\\
\hline
Adoption in Late June & 0.599 & 0.474\\
\hline
Adoption in May & 1.014 & 0.000\\
\hline
Adoption in June & 0.582 & 0.001\\
\hline
\end{tabular}
\caption{Effects estimation results for staggered adoption of treatment}
\label{tab:stag}

\end{table}

Looking at the results in Table \ref{tab:stag} and at the time series in Figure A1-A2 it can be seen that the biggest and most significant effect is in the states that started the policy early, along with the three states that started in the first half of June. In both cases, the policy seems to have induced an additional 1\% of the population to vaccinate against Covid-19. 

The other strata, which started the policy at the end of May and the end of June, show no significant effect. 

Focusing on aggregation at the month level, we notice that the treatment effect is positive in May and June, with a higher impact for the first month. Nevertheless, both results are significant at level $\alpha=0.05$.
Results suggest that there may have been a novelty effect due to the media attention given to this type of policy, which may have increased the effect of the policy in the first period. However, even if the June effect results are smaller than those in May, the positive sign of the effect should suggest an intrinsic positive effect from the policy, untied from the media attention.

\subsection{Treatment heterogeneity analysis}

In the \ref{sec:cluster} section, we have identified six clusters of treated counties, according to a matrix of socio-economic characteristics. We now use that classification to calculate the average treatment effect within those clusters.

We report the results of the estimation in the Table \ref{tab:ateclust}.

Looking at the results, we see that the treatment effect is positive in the majority of the counties. We observe a positive effect for clusters 1, 2, 5, and 6, which account for about 80\% of the treated units; cluster 4 results are positive but minimal. Specifically, clusters 1 and 6, markedly Republican, showed promising results from the policy, especially cluster 6, which was particularly affected by deaths due to Covid-19. Good results were also obtained in cluster 2, which tended to be wealthy, democratic and more educated on average, with persistence in the post-treatment. Negative or no results were obtained in clusters with more black and Hispanic ethnicity, accounting for 20\% of the total. From the results found, counties with higher ethnicity and lower wealth have worse policy outcomes. This result is consistent with previous findings in the literature, which found a link between ethnicity and vaccine hesitancy (\cite{quinn2016}, \cite{reiter2020}). This hesitancy appears to be unaffected by monetary incentives, in contrast with the idea that a monetary incentive can alter the choices of the less wealthy population.

In general, however, although differences in results are present, large standard errors of the average treatment effect suggest that none of these effects differs from zero.

\begin{table}[h]

\caption{ATE per cluster}

\centering

\footnotesize

\label{tab:ateclust}

\begin{tabular}[t]{|c|c|c|c|c|c|}

\hline

Cluster & Treatment & Post-treatment & sd treatment & sd post-treatment & \%\\

\hline

1 & 0.658 & 0.189 & 1.048 & 0.781 & 0.204\\

\hline

2 & 0.372 & 0.212 & 1.065 & 0.812 & 0.205\\

\hline

3 & -0.247 & -0.112 & 1.509 & 1.708 & 0.126\\

\hline

4 & 0.059 & -0.143 & 1.202 & 1.066 & 0.095\\

\hline

5 & 0.347 & -0.103 & 1.060 & 1.076 & 0.186\\

\hline

6 & 0.661 & 0.223 & 1.005 & 0.945 & 0.184\\

\hline

\end{tabular}

\end{table}
\FloatBarrier

\section{Conclusion}\label{sec:con}

 Policies to incentive vaccination through monetary transfers and lotteries have been rather discussed in the literature, finding possible benefits in the face of ethical and equity issues (\cite{jecker2021}, \cite{dotlic2021}, \cite{kim2021}, \cite{sprengholz2021}). This paper aims to study in-depth the vaccine lotteries promoted in different states of the United States, looking at the different results obtained at the county and state level and analyzing the role of timing in announcing such policies. 

 Conditional cash lotteries can be seen as a viable health policy to incentive people to get a vaccine jab, especially when the vaccination timing is a crucial variable in determining the outcome of the vaccination rollout. However, the experience of several US states, first introduced in Ohio in May 2021, suggests that the outcome of such policies is not unidirectional. In particular, we observed contrary results between different states and inside the same state, between different counties. 

 This applied study aims to explain the effects of incentives in the participating states, both at different levels of analysis (county, state, interstate) and at different points in time, providing an overall picture of policy outcomes. In particular, no previous study has focused on counties and the different characteristics of each unit. 

The results found were interesting and confirm the expected significant heterogeneity in treatment effects (\cite{dube2014} \cite{acharya2021}), affirming that there are differences in treatment effects between counties in the same states, suggesting that the social characteristics that determine this heterogeneity should be studied. Ohio, Washington, Kentucky, West Virginia, Illinois and Nevada outperformed other states obtaining positive and significant results, while the other states obtain no to negative results.
Furthermore, the analysis of staggered adoption effects suggests that 
states obtaining the best results were those that adopted the policy early, probably exploiting a media-buzz effect around the policy also hypothesized by \cite{gorin2015}. Nevertheless, positive results can still be obtained even with the late adoption of the policy.

The study of the effects after the end of the lottery is also unprecedented. This analysis is particularly relevant as it differentiates between counties and states that have experienced permanent effects from the policy or only temporary ones. This dimension might be decisive for the policymaker, who might only adopt this kind of incentive if a policy goal has to be achieved in a short period.  
We observed temporary positive effects in most counties treated, most pronounced in the Midwest and Northwestern areas. However, poor results were obtained in the Sunbelt, where monetary incentives were probably insufficient to convince a population vehemently opposed to the vaccine. 

In general, after an initial success, maybe backed up by the media attention, the vaccine lotteries did not significantly affect the vaccination choices of Americans, managing to involve someone who would probably have been vaccinated later.

\bibliographystyle{apalike}

\bibliography{bibtex_ref}

\section{Appendix}

\end{document}